\documentclass{article}
\usepackage{amsfonts,amssymb,amsmath,graphicx,hyperref}
\begin{document}

\title{New Mechanics of Spinal Injury}
\author{Vladimir G. Ivancevic}
\date{}
\maketitle

\begin{abstract}
The prediction and prevention of spinal injury is an important
aspect of preventive health science. The spine, or vertebral
column, represents a chain of 26 movable vertebral bodies, joint
together by transversal viscoelastic intervertebral discs and
longitudinal elastic tendons. This paper proposes a new
\emph{locally--coupled loading--rate hypothesis}, which states
that the main cause of both soft-- and hard--tissue spinal injury
is a \textsl{localized Euclidean jolt}, or $SE(3)-$jolt, an
impulsive loading that strikes a localized spine in several
coupled degrees-of-freedom simultaneously. To show this, based on
the previously defined \emph{covariant force law}, we formulate
the coupled Newton--Euler dynamics of the local spinal motions and
derive from it the corresponding coupled $SE(3)-$jolt dynamics. The $SE(3)-$%
jolt is the main cause of two basic forms of spinal injury: (i)
hard--tissue injury of local translational dislocations; and (ii)
soft--tissue injury of local rotational disclinations. Both the
spinal \emph{dislocations and disclinations}, as caused by the
$SE(3)-$jolt, are described using the Cosserat multipolar
viscoelastic continuum model. \bigbreak

dynamics,
Euclidean jolt dynamics, spinal dislocations and disclinations
\end{abstract}

\vspace{5cm}

\noindent\textbf{Contact information:}\bigbreak

\noindent Dr. Vladimir Ivancevic\newline
Human Systems Integration, Land Operations Division\newline
Defence Science \& Technology Organisation, AUSTRALIA\newline
PO Box 1500, 75 Labs, Edinburgh SA 5111\newline
Tel:~ +61 8 8259 7337, ~~~Fax:~ +61 8 8259 4193\newline
E-mail: ~ Vladimir.Ivancevic$@$dsto.defence.gov.au

%\tableofcontents

\newpage

\section{Introduction}

Normal function of the human spine is possible due to a complex
interaction of its components (i.e., vertebrae, ligaments, discs,
rib cage, and muscles). Age, trauma, spinal disorders, and a host
of other parameters can disrupt this interaction to an extent that
in certain cases surgery may be required to restore normal
function. Several spinal disorders have been described in
\cite{Goel} from a mechanical perspective. An understanding of
these disorders can assist in the design and development of spinal
instrumentation. As biomechanics begins to be intertwined with
tissue engineering, a better understanding of the particular
disorders may also provide insight into `biological' solutions.

In particular, the center of rotation of the upper cervical spine
is an important biomechanical landmark that is used to determine
upper neck moment, particularly when evaluating injury risk in the
automotive environment \cite{Chancey}. Also, new vehicle safety
standards are designed to limit the amount of neck tension and
extension seen by out-of-position motor vehicle occupants during
airbag deployments. The criteria used to assess airbag injury risk
are currently based on volunteer data and animal studies due to a
lack of bending tolerance data for the adult cervical spine
\cite{Nightingale}.

Also, lumbar spine pathology accounts for billions of dollars in
societal costs each year. Although the symptomatology of these
conditions is relatively well understood, the mechanical changes
in the spine are not. Previous direct measurements of lumbar spine
mechanics have mostly been performed on cadavers. The methods for
in vivo studies have included imaging, electrogoniometry, and
motion capture. Few studies have directly measured in vivo lumbar
spine kinematics with in-dwelling bone pins. In vivo 3D motion of
the entire lumbar spine has recently been tracked during gait in
\cite{Rozumalski}. Using a direct (pin-based) in vivo measurement
method, the motion of the human lumbar spine during gait was found
to be triaxial. This appears to be the first 3D motion analysis of
the entire lumbar spine using indwelling pins. The results were
similar to previously published data derived from a variety of
experimental methods.

The traditional \textit{principal loading hypothesis}
\cite{McElhaney,Whiting}, which describes general spinal injuries
in terms of spinal tension, compression, bending, and shear, is
insufficient to predict and prevent the cause of the back-pain
syndrome. Its underlying mechanics is simply not accurate enough.
On the other hand, to be recurrent, musculo-skeletal injury must
be associated with a histological change, i.e., the modification
of associated tissues within the body. However, incidences of
\textit{functional} musculoskeletal injury, e.g., lower back pain,
generally shows little evidence of \textit{structural} damage
\cite{Waddell}. The incidence of injury is likely to be a
continuum ranging from little or no evidence of structural damage
through to the observable damage of muscles, joints or bones. The
changes underlying functional injuries are likely to consist of
torn muscle fibers, stretched ligaments, subtle erosion of join
tissues, and/or the application of pressure to nerves, all
amounting to a disruption of function to varying degrees and a
tendency toward spasm.

For example, in a review of experimental studies on the role of
mechanical stresses in the genesis of intervertebral disk
degeneration and herniation \cite{Rannou}, the authors dismissed
simple mechanical stimulations of functional vertebra as a cause
of disk herniation, concluding instead that a complex mechanical
stimulation combining forward and lateral bending of the spine
followed by violent compression is needed to produce posterior
herniation of the disk. Considering the use of models to estimate
the risk of injury the authors emphasize the need to understand
this complex interaction between the mechanical forces and the
living body \cite{Seidel}. Compressive and shear loading increased
significantly with exertion load, lifting velocity, and trunk
asymmetry \cite{Granata}. Also, it has been stated that up to
two--thirds of all back injuries have been associated with trunk
rotation \cite{Kumar}. In addition, load--lifting in awkward
environment places a person at risk for low back pain and injury
\cite{Reiser}. These risks appear to be increased when facing up
or down an inclined surface.

The safe spinal motions (flexion/extension, lateral flexion and
rotation) \textit{are} governed by standard Euler's rotational
intervertebral dynamics coupled to Newton's micro-translational
dynamics. On the other hand, the unsafe spinal events, the main
cause of spinal injuries, are caused by intervertebral
SE(3)--jolts, the sharp and sudden, \textquotedblleft
delta\textquotedblright -- (forces + torques) combined, localized
both in time and in space. These localized intervertebral
SE(3)--jolts do not belong to the standard Newton--Euler dynamics.
The only way to monitor them would be to measure ``in vivo" the
rate of the combined (forces + torques)-- rise.

It is well known that the mechanical properties of spinal
ligaments and muscles are rate dependent. As elongation rate
increases, ligaments generally exhibit higher stiffness, higher
failure force, and smaller failure strain. Previous studies have
shown that high-speed multiplanar loading causes soft tissue
injury that is more severe as compared to sagittal loading. This
paper proposes a new {locally--coupled loading--rate hypothesis},
which states that the main cause of both soft-- and hard--tissue
spinal injury is a {localized Euclidean jolt}, or $SE(3)-$jolt, an
impulsive loading that strikes a localized spine in several
coupled degrees-of-freedom (DOF) simultaneously. To show this,
based on the previously defined {covariant force law}, we
formulate the coupled Newton--Euler dynamics of the local spinal
motions and derive from it the corresponding coupled $SE(3)-$jolt
dynamics. The $SE(3)-$jolt is the main cause of two forms of local
discontinuous spinal injury: (i) hard--tissue injury of local
translational dislocations; and (ii) soft--tissue injury of local
rotational disclinations. Both the spinal {dislocations and
disclinations}, as caused by the $SE(3)-$jolt, are described using
the Cosserat multipolar viscoelastic continuum model.

While we can intuitively visualize the SE(3)--jolt, for the
purpose of simulation we use the necessary simplified, decoupled
approach (neglecting the 3D torque matrix and its coupling to the
3D force vector). Note that decoupling is a kind of linearization
that prevents chaotic behavior, giving an illusion of full
predictability. In this decoupled framework of reduced complexity,
we define:

The cause of hard spinal injuries (discus hernia) is a linear
3D--jolt vector hitting some intervertebral joint -- the time
rate-of-change of a 3D--force vector (linear jolt = mass $\times $
linear jerk).

The cause of soft spinal injuries (back--pain syndrome) is an
angular 3--axial jolt hitting some intervertebral joint -- the
time rate-of-change of a 3--axial torque (angular jolt = inertia
moment $\times $ angular jerk).

This decoupled framework has been implemented in the Human
Biodynamics Engine \cite{18}, a world--class
neuro--musculo--skeletal dynamics simulator (with 270 DOFs, the
same number of equivalent muscular actuators and two--level neural
reflex control), developed by the present author at Defence
Science and Technology Organization, Australia. This kinematically
validated human motion simulator has been described in a series of
papers and books \cite{6,7,8,9,11},\\
\cite{12,13,14},\\
\cite{17,IJHR,NeuFuz,StrAttr,CompMind,Complexity}.

\section{The $SE(3)-$jolt: the main cause of spinal injury}

\begin{figure}[tbp]
\centering \includegraphics[width=14cm]{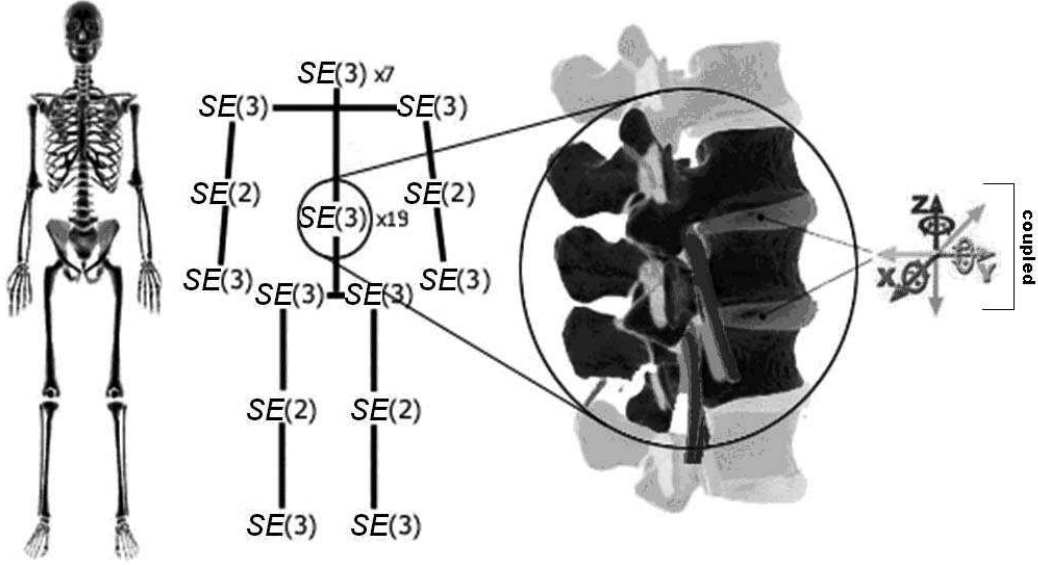} \caption{Human
body representation in terms of SE(3)/SE(2)--groups of rigid--body
motion, with the vertebral column represented as a chain of 26
flexibly--coupled SE(3)--groups.} \label{IvSpine}
\end{figure}

In the language of modern biodynamics \cite{9,12},\newline
\cite{13,14,14a},\newline \cite{17}, the general spinal motion is
governed by the Euclidean SE(3)--group of 3D motions (see Figure
\ref{IvSpine}). Within the spinal SE(3)--group we have both
SE(3)--kinematics (consisting of the spinal SE(3)--velocity and
its two time derivatives: SE(3)--acceleration and SE(3)--jerk) and
the spinal SE(3)--dynamics (consisting of SE(3)--momentum and its
two time derivatives: SE(3)--force and SE(3)--jolt), which is the
spinal kinematics $\times $ the spinal mass--inertia distribution.

Informally, the \textit{localized spinal SE(3)--jolt}\footnote{%
The mechanical SE(3)--jolt concept is based on the mathematical concept of
higher--order tangency (rigorously defined in terms of jet bundles of the
head's configuration manifold) \cite{14,17}, as follows: When something hits
the human head, or the head hits some external body, we have a collision.
This is naturally described by the SE(3)--momentum, which is a nonlinear
coupling of 3 linear Newtonian momenta with 3 angular Eulerian momenta. The
tangent to the SE(3)--momentum, defined by the (absolute) time derivative,
is the SE(3)--force. The second-order tangency is given by the SE(3)--jolt,
which is the tangent to the SE(3)--force, also defined by the time
derivative.} is a sharp and sudden change in the localized spinal
SE(3)--force acting on the localized spinal mass--inertia distribution. That
is, a `delta'--change in a 3D force--vector coupled to a 3D torque--vector,
striking the certain local point along the vertebral column. In other words,
the localized spinal SE(3)--jolt is a sudden, sharp and discontinues shock
in all 6 coupled dimensions of a local spinal point, within the three
Cartesian ($x,y,z$)--translations and the three corresponding Euler angles
around the Cartesian axes: roll, pitch and yaw \cite{7}. If the SE(3)--jolt
produces a mild shock to the spine, it causes mild, soft--tissue spinal
injury, usually resulting in the back--pain sindrome. If the SE(3)--jolt
produces a hard shock to the spine, it causes severe, hard--tissue spinal
injury, with the total loss of movement.

Therefore, we propose a new \emph{combined loading--rate hypothesis} of the
local spinal injury instead of the old principal loading hypothesis. This
new hypothesis has actually been supported by a number of individual
studies, both experimental and numerical, as can be seen from the following
brief review. One of the first dynamical studies of the head--neck system's
response to impulsive loading was performed in \cite{Misra2}. The response
of a human head/neck/torso system to shock was investigated in \cite{Luo91},
using a 3D numerical and physical models; the results indicated that the
head, cervical muscles and disks in the lumbar region were subjected to the
greatest \emph{force changes} and thus were most likely to be injured.
Time--dependent changes in the lumbar spine's resistance to bending was
investigated in \cite{Adams96}, with the objective to show how time--related
factors might affect the risk of back injury; the results suggested that the
risk of bending injury to the lumbar discs and ligaments would depend not
only on the loads applied to the spine, but also on \emph{loading rate}.
Cyclic loading tests were performed by \cite{Tsai98} to investigate the
mechanical responses at different loading rates; the results indicated that
faster \emph{loading rate} generated greater stress decay, and disc
herniation was more likely to occur under higher loading rate conditions.
Anterior shear of spinal motion segments was experimentally investigated in
\cite{Yingling99}; kinematics, kinetics, and resultant injuries were
observed; dynamic loading and flexion of the specimens were found to
increase the ultimate load at failure when compared with quasi-static
loading and neutral postures. Experimental evidence concerning the
distribution of forces and moments acting on the lumbar spine was reviewed
in \cite{Dolan01}, pointing out that it was necessary to distribute the
overall forces and moments between (and within) different spinal structures,
because it was the \emph{concentration of force} which caused injury, and
elicited pain. Small magnitudes of axial torque was shown to in \cite%
{Drake05} to alter the failure mechanics of the intervertebral disc and
vertebrae in \emph{combined loading} situations. A finite element model of
head and cervical spine based on the actual geometry of a human cadaver
specimen was developed in \cite{Zhang06}, which predicted the \emph{%
nonlinear moment-rotation relationship} of human cervical spine. Vertebral
end-plate fractures as a result of high--rate pressure loading were
investigated in \cite{Brown08}, where a slightly exponential relationship
was found between \emph{peak pressure} and its \emph{rate of development}.

The localized spinal SE(3)--jolt is rigorously defined in terms of
differential geometry\\ \cite{14,17}. Briefly, it is the absolute
time--derivative of the covariant force 1--form (or, co-vector
field) applied to the spine at a certain local point. With this
respect, recall that the fundamental law of biomechanics -- the
so--called \emph{covariant force law} \cite{13,14,17}, states:
\begin{equation*}
\text{Force co-vector field}=\text{Mass distribution}\times \text{%
Acceleration vector--field},
\end{equation*}%
which is formally written (using the Einstein summation convention, with
indices labelling the three local Cartesian translations and the
corresponding three local Euler angles):
\begin{equation*}
F_{{\mu }}=m_{{\mu }{\nu }}a^{{\nu }},\qquad ({\mu ,\nu }=1,...,6=3\text{
Cartesian}+3\text{ Euler})
\end{equation*}%
where $F_{{\mu }}$ denotes the 6 covariant components of the localized
spinal SE(3)--force co-vector field, $m_{{\mu }{\nu }}$ represents the 6$%
\times $6 covariant components of the localized spinal inertia--metric
tensor, while $a^{{\nu }}$ corresponds to the 6 contravariant components of
localized spinal SE(3)--acceleration vector-field.

Now, the covariant (absolute, Bianchi) time--derivative $\frac{{D}}{dt}%
(\cdot )$ of the covariant SE(3)--force $F_{{\mu }}$ defines the
corresponding localized spinal SE(3)--jolt co-vector field:
\begin{equation}
\frac{{D}}{dt}(F_{{\mu }})=m_{{\mu }{\nu }}\frac{{D}}{dt}(a^{{\nu }})=m_{{%
\mu }{\nu }}\left( \dot{a}^{{\nu }}+\Gamma _{\mu \lambda }^{{\nu }}a^{{\mu }%
}a^{{\lambda }}\right) ,  \label{Bianchi}
\end{equation}%
where ${\frac{{D}}{dt}}{(}a^{{\nu }})$ denotes the 6 contravariant
components of the localized spinal SE(3)--jerk vector-field and overdot ($%
\dot{~}$) denotes the time derivative. $\Gamma _{\mu \lambda
}^{{\nu }}$ are the Christoffel's symbols of the Levi--Civita
connection for the SE(3)--group, which are zero in case of pure
Cartesian translations and nonzero in case of rotations as well as
in the full--coupling of translations and rotations.

In the following, we elaborate on the localized spinal SE(3)--jolt concept
(using vector and tensor methods) and its biophysical consequences in the
form of the localized spinal dislocations and disclinations.

\subsection{$SE(3)-$group of local spinal motions}

Briefly, the $SE(3)-$group of localized spinal motions is defined as a
semidirect (noncommutative) product of 3D intervertebral rotations and 3D
intervertebral micro--translations,
\begin{equation*}
SE(3):=SO(3)\rhd \mathbb{R}^{3}.
\end{equation*}%
Its most important subgroups are the following (see Appendix for technical
details):

\begin{center}
{{\frame{$
\begin{array}{cc}
\mathbf{Subgroup} & \mathbf{Definition} \\ \hline
\begin{array}{c}
SO(3),\text{ group of rotations} \\
\text{in 3D (a spherical joint)}%
\end{array}
&
\begin{array}{c}
\text{Set of all proper orthogonal } \\
3\times 3-\text{rotational matrices}%
\end{array}
\\ \hline
\begin{array}{c}
SE(2),\text{ special Euclidean group} \\
\text{in 2D (all planar motions)}%
\end{array}
&
\begin{array}{c}
\text{Set of all }3\times 3-\text{matrices:} \\
\left[
\begin{array}{ccc}
\cos \theta & \sin \theta & r_{x} \\
-\sin \theta & \cos \theta & r_{y} \\
0 & 0 & 1%
\end{array}
\right]%
\end{array}
\\ \hline
\begin{array}{c}
SO(2),\text{ group of rotations in 2D} \\
\text{subgroup of }SE(2)\text{--group} \\
\text{(a revolute joint)}%
\end{array}
&
\begin{array}{c}
\text{Set of all proper orthogonal } \\
2\times 2-\text{rotational matrices} \\
\text{ included in }SE(2)-\text{group}%
\end{array}
\\ \hline
\begin{array}{c}
\mathbb{R}^{3},\text{ group of translations in 3D} \\
\text{(all spatial displacements)}%
\end{array}
& \text{Euclidean 3D vector space}%
\end{array}
$}}}
\end{center}

In other words, the gauge $SE(3)-$group of intervertebral Euclidean
micro-motions contains matrices of the form {\small $\left(
\begin{array}{cc}
\mathbf{R} & \mathbf{p} \\
0 & 1%
\end{array}%
\right) ,$} where $\mathbf{p}$ is intervertebral 3D micro-translation vector
and $\mathbf{R}$ is intervertebral 3D rotation matrix, given by the product $%
\mathbf{R}=R_{\varphi }\cdot R_{\psi }\cdot R_{\theta }$ of the three
Eulerian intervertebral rotations, $\text{roll}=R_{\varphi },~\text{pitch}%
=R_{\psi },~\text{yaw}=R_{\theta }$, performed respectively about the $x-$%
axis by an angle $\varphi ,$ about the $y-$axis by an angle $\psi ,$ and
about the $z-$axis by an angle $\theta $ (see \cite{9,ParkChung,IJHR}%
), {\small
\begin{equation*}
R_{\varphi }=\left[
\begin{array}{ccc}
1 & 0 & 0 \\
0 & \cos \varphi & -\sin \varphi \\
0 & \sin \varphi & \cos \varphi%
\end{array}%
\right] ,~~R_{\psi }=\left[
\begin{array}{ccc}
\cos \psi & 0 & \sin \psi \\
0 & 1 & 0 \\
-\sin \psi & 0 & \cos \psi%
\end{array}%
\right] ,~~R_{\theta }=\left[
\begin{array}{ccc}
\cos \theta & -\sin \theta & 0 \\
\sin \theta & \cos \theta & 0 \\
0 & 0 & 1%
\end{array}%
\right] .
\end{equation*}%
}

Therefore, natural intervertebral $SE(3)-$dynamics is given by the coupling
of Newtonian (translational) and Eulerian (rotational) equations of
intervertebral motion.

\subsection{Localized spinal $SE(3)-$dynamics}

To support our locally--coupled loading--rate hypothesis, we formulate the
coupled Newton--Euler dynamics of localized spinal motions within the $%
SE(3)- $group. The forced Newton--Euler equations read in vector (boldface)
form
\begin{eqnarray}
\text{Newton} &:&~\mathbf{\dot{p}}~\mathbf{\equiv M\dot{v}=F+p\times \omega }%
,  \label{vecForm} \\
\text{Euler} &:&~\mathbf{\dot{\pi}}~\mathbf{\equiv I\dot{\omega}=T+\pi
\times \omega +p\times v},  \notag
\end{eqnarray}%
where $\times $ denotes the vector cross product,\footnote{%
Recall that the cross product $\mathbf{u\times v}$ of two vectors $\mathbf{u}
$ and $\mathbf{v}$ equals $\mathbf{u\times v}=uv{\rm sin}\theta \mathbf{n}$%
, where $\theta $ is the angle between $\mathbf{u}$ and $\mathbf{v}$, while $%
\mathbf{n}$ is a unit vector perpendicular to the plane of $\mathbf{u}$ and $%
\mathbf{v}$ such that $\mathbf{u}$ and $\mathbf{v}$ form a right-handed
system.}
\begin{equation*}
\mathbf{M}\equiv M_{ij}=diag\{m_{1},m_{2},m_{3}\}\qquad \text{and}\qquad
\mathbf{I}\equiv I_{ij}=diag\{I_{1},I_{2},I_{3}\},\qquad (i,j=1,2,3)
\end{equation*}%
are spinal segment's (diagonal) mass and inertia matrices,\footnote{%
In reality, mass and inertia matrices ($\mathbf{M,I}$) are not diagonal but
rather full $3\times 3$ positive--definite symmetric matrices with coupled
mass-- and inertia--products. Even more realistic, fully--coupled
mass--inertial properties of a spinal segment are defined by the single
non-diagonal $6\times 6$ positive--definite symmetric mass--inertia matrix $%
\mathcal{M}_{SE(3)}$, the so-called material metric tensor of the $SE(3)-$%
group, which has all nonzero mass--inertia coupling products. However, for
simplicity, in this paper we shall consider only the simple case of two
separate diagonal $3\times 3$ matrices ($\mathbf{M,I}$).} defining the
localized spinal mass--inertia distribution, with principal inertia moments
given in Cartesian coordinates ($x,y,z$) by volume integrals
\begin{equation*}
I_{1}=\iiint \rho (z^{2}+y^{2})dxdydz,~~I_{2}=\iiint \rho
(x^{2}+z^{2})dxdydz,~~I_{3}=\iiint \rho (x^{2}+y^{2})dxdydz,
\end{equation*}%
dependent on localized spinal density $\rho =\rho (x,y,z)$,
\begin{equation*}
\mathbf{v}\equiv v^{i}=[v_{1},v_{2},v_{3}]^{t}\qquad \text{and\qquad }%
\mathbf{\omega }\equiv {\omega }^{i}=[\omega _{1},\omega _{2},\omega
_{3}]^{t}
\end{equation*}%
(where $[~]^{t}$ denotes the vector transpose) are localized spinal linear
and angular velocity vectors\footnote{%
In reality, $\mathbf{\omega }$ is a $3\times 3$ \emph{attitude matrix} (see
Appendix). However, for simplicity, we will stick to the (mostly)
symmetrical translation--rotation vector form.} (that is, column vectors),
\begin{equation*}
\mathbf{F}\equiv F_{i}=[F_{1},F_{2},F_{3}]\qquad \text{and}\qquad \mathbf{T}%
\equiv T_{i}=[T_{1},T_{2},T_{3}]
\end{equation*}%
are gravitational and other external force and torque co-vectors (that is,
row vectors) acting on the spine,
\begin{eqnarray*}
\mathbf{p} &\equiv &p_{i}\equiv \mathbf{Mv}%
=[p_{1},p_{2},p_{3}]=[m_{1}v_{1},m_{2}v_{2},m_{2}v_{2}]\qquad \text{and} \\
\mathbf{\pi } &\equiv &\pi _{i}\equiv \mathbf{I\omega }=[\pi _{1},\pi
_{2},\pi _{3}]=[I_{1}\omega _{1},I_{2}\omega _{2},I_{3}\omega _{3}]
\end{eqnarray*}%
are localized spinal linear and angular momentum co-vectors.

In tensor form, the forced Newton--Euler equations (\ref{vecForm}) read
\begin{eqnarray*}
\dot{p}_{i} &\equiv &M_{ij}\dot{v}^{j}=F_{i}+\varepsilon _{ik}^{j}p_{j}{%
\omega }^{k},\qquad(i,j,k=1,2,3) \\
\dot{\pi}_{i} &\equiv &I_{ij}\dot{\omega}^{j}=T_{i}+\varepsilon _{ik}^{j}\pi
_{j}\omega ^{k}+\varepsilon _{ik}^{j}p_{j}v^{k},
\end{eqnarray*}
where the permutation symbol $\varepsilon _{ik}^{j}$ is\ defined as
\begin{equation*}
\varepsilon _{ik}^{j}=
\begin{cases}
+1 & \text{if }(i,j,k)\text{ is }(1,2,3),(3,1,2)\text{ or }(2,3,1), \\
-1 & \text{if }(i,j,k)\text{ is }(3,2,1),(1,3,2)\text{ or }(2,1,3), \\
0 & \text{otherwise: }i=j\text{ or }j=k\text{ or }k=i.%
\end{cases}%
\end{equation*}

In scalar form, the forced Newton--Euler equations (\ref{vecForm}) expand as
\begin{eqnarray}
\text{Newton} &:&\left\{
\begin{array}{c}
\dot{p}_{_{1}}={F_{1}}-{m_{3}}{v_{3}}{\omega _{2}}+{m_{2}}{v_{2}}{\omega _{3}%
} \\
\dot{p}_{_{2}}={F_{2}}+{m_{3}}{v_{3}}{\omega _{1}}-{m_{1}}{v_{1}}{\omega _{3}%
} \\
\dot{p}_{_{3}}={F_{3}}-{m_{2}}{v_{2}}{\omega _{1}}+{m_{1}}{v_{1}}{\omega _{2}%
}%
\end{array}%
\right. ,  \label{scalarForm} \\
\text{Euler} &:&\left\{
\begin{array}{c}
\dot{\pi}_{_{1}}={T_{1}}+({m_{2}}-{m_{3}}){v_{2}}{v_{3}}+({I_{2}}-{I_{3}}){%
\omega _{2}}{\omega _{3}} \\
\dot{\pi}_{_{2}}={T_{2}}+({m_{3}}-{m_{1}}){v_{1}}{v_{3}}+({I_{3}}-{I_{1}}){%
\omega _{1}}{\omega _{3}} \\
\dot{\pi}_{_{3}}={T_{3}}+({m_{1}}-{m_{2}}){v_{1}}{v_{2}}+({I_{1}}-{I_{2}}){%
\omega _{1}}{\omega _{2}}%
\end{array}%
\right. ,  \notag
\end{eqnarray}%
showing localized spinal mass and inertia couplings.

Equations (\ref{vecForm})--(\ref{scalarForm}) can be derived from the
translational + rotational kinetic energy of the spine segment\footnote{%
In a fully--coupled Newton--Euler localized spinal dynamics, instead of
equation (\ref{Ek}) we would have spinal segment's kinetic energy defined by
the inner product:
\begin{equation*}
E_{k}=\frac{1}{2}\left[{\mathbf{p}}{\mathbf{\pi }}\left\vert
\mathcal{M}_{SE(3)}\right.{\mathbf{p}}{\mathbf{\pi }}\right] .
\end{equation*}%
}
\begin{equation}
E_{k}={\frac{1}{2}}\mathbf{v}^{t}\mathbf{Mv}+{\frac{1}{2}}\mathbf{\omega }%
^{t}\mathbf{I\omega },  \label{Ek}
\end{equation}%
or, in tensor form
\begin{equation*}
E={\frac{1}{2}}M_{ij}{v}^{i}{v}^{j}+{\frac{1}{2}}I_{ij}{\omega}%
^{i}{\omega}^{j}.
\end{equation*}

For this we use the \emph{Kirchhoff--Lagrangian equations} (see, e.g., \cite%
{Kirchhoff,naomi97}, or the original work of Kirchhoff in German)
\begin{eqnarray}
\frac{d}{{dt}}\partial _{\mathbf{v}}E_{k} &=&\partial _{\mathbf{v}%
}E_{k}\times \mathbf{\omega }+\mathbf{F},  \label{Kirch} \\
{\frac{d}{{dt}}}\partial _{\mathbf{\omega }}E_{k} &=&\partial _{\mathbf{%
\omega }}E_{k}\times \mathbf{\omega }+\partial _{\mathbf{v}}E_{k}\times
\mathbf{v}+\mathbf{T},  \notag
\end{eqnarray}
where $\partial _{\mathbf{v}}E_{k}=\frac{\partial E_{k}}{\partial \mathbf{v}}%
,~\partial _{\mathbf{\omega }}E_{k}=\frac{\partial E_{k}}{\partial \mathbf{%
\omega }}$; in tensor form these equations read
\begin{eqnarray*}
\frac{d}{dt}\partial _{v^{i}}E &=&\varepsilon _{ik}^{j}\left( \partial
_{v^{j}}E\right) \omega ^{k}+F_{i}, \\
\frac{d}{dt}\partial _{{\omega }^{i}}E &=&\varepsilon _{ik}^{j}\left(
\partial _{{\omega }^{j}}E\right) {\omega }^{k}+\varepsilon _{ik}^{j}\left(
\partial _{v^{j}}E\right) v^{k}+T_{i}.
\end{eqnarray*}

Using (\ref{Ek})--(\ref{Kirch}), localized spinal linear and angular
momentum co-vectors are defined as
\begin{equation*}
\mathbf{p}=\partial _{\mathbf{v}}E_{k}{,\qquad \mathbf{\pi }=\partial _{%
\mathbf{\omega }}E_{k},}
\end{equation*}%
or, in tensor form
\begin{equation*}
p_{i}=\partial _{v^{i}}E{,\qquad }\pi _{i}=\partial _{{\omega }^{i}}E,
\end{equation*}%
with their corresponding time derivatives, in vector form
\begin{equation*}
~\mathbf{\dot{p}}=\frac{d}{dt}\mathbf{p=}\frac{d}{dt}\partial _{\mathbf{v}}E{%
,\qquad \mathbf{\dot{\pi}}=}\frac{d}{dt}\mathbf{\pi =}\frac{d}{dt}\partial _{%
\mathbf{\omega }}E,
\end{equation*}%
or, in tensor form
\begin{equation*}
~\dot{p}_{i}=\frac{d}{dt}p_{i}=\frac{d}{dt}\partial _{v^{i}}E{,\qquad \dot{%
\pi}_{i}=}\frac{d}{dt}\pi _{i}=\frac{d}{dt}\partial _{{\omega }^{i}}E,
\end{equation*}%
or, in scalar form
\begin{equation*}
\mathbf{\dot{p}}=[\dot{p}_{1},\dot{p}_{2},\dot{p}_{3}]=[m_{1}\dot{v}%
_{1},m_{2}\dot{v}_{2},m_{3}\dot{v}_{3}],\qquad {\mathbf{\dot{\pi}}}=[\dot{\pi%
}_{1},\dot{\pi}_{2},\dot{\pi}_{3}]=[I_{1}\dot{\omega}_{1},I_{2}\dot{\omega}%
_{2},I_{3}\dot{\omega}_{3}].
\end{equation*}

While spinal healthy $SE(3)-$dynamics is given by the coupled Newton--Euler
micro--dynamics, the localized spinal injury is actually caused by the sharp
and discontinuous change in this natural $SE(3)$ micro-dynamics, in the form
of the $SE(3)-$jolt, causing localized discontinuous spinal deformations,
both translational dislocations and rotational disclinations.

\subsection{Localized spinal--injury dynamics: the $SE(3)-$jolt}

The $SE(3)-$jolt, the actual cause of spinal injury (in the form
of the localized spinal plastic deformations), is defined as a
coupled Newton+Euler jolt; in
(co)vector form the $SE(3)-$jolt reads\footnote{%
Note that the derivative of the cross--product of two vectors follows the
standard calculus product--rule: $\frac{d}{dt}(\mathbf{u\times v})=\mathbf{%
\dot{u}\times v+u\times \dot{v}.}$}
\begin{equation*}
SE(3)-\text{jolt}:\left\{
\begin{array}{l}
\text{Newton~jolt}:\mathbf{\dot{F}=\ddot{p}-\dot{p}\times \omega -p\times
\dot{\omega}}~,\qquad \\
\text{Euler~jolt}:\mathbf{\dot{T}=\ddot{\pi}}~\mathbf{-\dot{\pi}\times
\omega -\pi \times \dot{\omega}-\dot{p}\times v-p\times \dot{v}},%
\end{array}
\right.
\end{equation*}
where the linear and angular jolt co-vectors are
\begin{equation*}
\mathbf{\dot{F}\equiv M\ddot{v}}=[\dot{F}_{{1}},\dot{F}_{{2}},\dot{F}_{{3}%
}],\qquad \mathbf{\dot{T}\equiv I\ddot{\omega}}=[\dot{T}_{{1}},\dot{T}_{{2}},%
\dot{T}_{{3}}],
\end{equation*}
where
\begin{equation*}
\mathbf{\ddot{v}}=[\ddot{v}_{{1}},\ddot{v}_{{2}},\ddot{v}_{{3}}]^{t},\qquad
\mathbf{\ddot{\omega}}=[\ddot{\omega}_{{1}},\ddot{\omega}_{{2}},\ddot{\omega}%
_{{3}}]^{t},
\end{equation*}
are linear and angular jerk vectors.

In tensor form, the $SE(3)-$jolt reads\footnote{%
In this paragraph the overdots actually denote the absolute Bianchi
(covariant) time-derivative (\ref{Bianchi}), so that the jolts retain the
proper covector character, which would be lost if ordinary time derivatives
are used. However, for the sake of simplicity and wider readability, we
stick to the same overdot notation.}
\begin{eqnarray*}
~\dot{F}_{i} &=&\ddot{p}_{i}-\varepsilon _{ik}^{j}\dot{p}_{j}{\omega }%
^{k}-\varepsilon _{ik}^{j}p_{j}{\dot{\omega}}^{k}, \qquad(i,j,k=1,2,3) \\
~\dot{T}_{{i}} &=&\ddot{\pi}_{i}~-\varepsilon _{ik}^{j}\dot{\pi}_{j}\omega
^{k}-\varepsilon _{ik}^{j}\pi _{j}{\dot{\omega}}^{k}-\varepsilon _{ik}^{j}%
\dot{p}_{j}v^{k}-\varepsilon _{ik}^{j}p_{j}\dot{v}^{k},
\end{eqnarray*}
in which the linear and angular jolt covectors are defined as
\begin{eqnarray*}
\mathbf{\dot{F}} &\equiv &\dot{F}_{i}=\mathbf{M\ddot{v}}\,\equiv \mathbf{\,}%
M_{ij}\ddot{v}^{j}=[\dot{F}_{1},\dot{F}_{2},\dot{F}_{3}], \\
\mathbf{\dot{T}} &\equiv &\dot{T}_{{i}}=\mathbf{I\ddot{\omega}\equiv \,}%
I_{ij}\ddot{\omega}^{j}=[\dot{T}_{{1}},\dot{T}_{{2}},\dot{T}_{{3}}],
\end{eqnarray*}
where \ $\mathbf{\ddot{v}}=\ddot{v}^{{i}},$ and $\mathbf{\ddot{\omega}}=%
\ddot{\omega}^{{i}}$ are linear and angular jerk vectors.

In scalar form, the $SE(3)-$jolt expands as
\begin{eqnarray*}
\text{Newton~jolt} &:&\left\{
\begin{array}{l}
\dot{F}_{{1}}=\ddot{p}_{1}-m_{{2}}\omega _{{3}}\dot{v}_{{2}}+m_{{3}}\left( {%
\omega }_{{2}}\dot{v}_{{3}}+v_{{3}}\dot{\omega}_{{2}}\right) -m_{{2}}v_{{2}}{%
\dot{\omega}}_{{3}}, \\
\dot{F}_{{2}}=\ddot{p}_{2}+m_{{1}}\omega _{{3}}\dot{v}_{{1}}-m_{{3}}\omega _{%
{1}}\dot{v}_{{3}}-m_{{3}}v_{{3}}\dot{\omega}_{{1}}+m_{{1}}v_{{1}}\dot{\omega}%
_{{3}}, \\
\dot{F}_{{3}}=\ddot{p}_{3}-m_{{1}}\omega _{{2}}\dot{v}_{{1}}+m_{{2}}\omega _{%
{1}}\dot{v}_{{2}}-v_{{2}}\dot{\omega}_{{1}}-m_{{1}}v_{{1}}\dot{\omega}_{{2}},%
\end{array}
\right. \\
&& \\
\text{Euler~jolt} &:&\left\{
\begin{array}{l}
\dot{T}_{{1}}=\ddot{\pi}_{1}-(m_{{2}}-m_{{3}})\left( v_{{3}}\dot{v}_{{2}}+v_{%
{2}}\dot{v}_{{3}}\right) -(I_{{2}}-I_{{3}})\left( \omega _{{3}}\dot{\omega}_{%
{2}}+{\omega }_{{2}}{\dot{\omega}}_{{3}}\right) , \\
\dot{T}_{{2}}=\ddot{\pi}_{2}+(m_{{1}}-m_{{3}})\left( v_{{3}}\dot{v}_{{1}}+v_{%
{1}}\dot{v}_{{3}}\right) +(I_{{1}}-I_{{3}})\left( {\omega }_{{3}}{\dot{\omega%
}}_{{1}}+{\omega }_{{1}}{\dot{\omega}}_{{3}}\right) , \\
\dot{T}_{{3}}=\ddot{\pi}_{3}-(m_{{1}}-m_{{2}})\left( v_{{2}}\dot{v}_{{1}}+v_{%
{1}}\dot{v}_{{2}}\right) -(I_{{1}}-I_{{2}})\left( {\omega }_{{2}}{\dot{\omega%
}}_{{1}}+{\omega }_{{1}}{\dot{\omega}}_{{2}}\right).%
\end{array}
\right.
\end{eqnarray*}

We remark here that the linear and angular momenta ($\mathbf{p,\pi }$),
forces ($\mathbf{F,T}$) and jolts ($\mathbf{\dot{F},\dot{T}}$) are
co-vectors (row vectors), while the linear and angular velocities ($\mathbf{%
v,\omega }$), accelerations ($\mathbf{\dot{v},\dot{\omega}}$) and jerks ($%
\mathbf{\ddot{v},\ddot{\omega}}$) are vectors (column vectors). This
bio-physically means that the `jerk' vector should not be confused with the
`jolt' co-vector. For example, the `jerk'\ means shaking the head's own
mass--inertia matrices (mainly in the atlanto--occipital and atlanto--axial
joints), while the `jolt'means actually hitting the head with some external
mass--inertia matrices included in the `hitting'\ SE(3)--jolt, or hitting
some external static/massive body with the head (e.g., the ground --
gravitational effect, or the wall -- inertial effect). Consequently, the
mass-less `jerk' vector\ represents a (translational+rotational) \textit{%
non-collision effect} that can cause only soft--tissue spinal injuries,
while the inertial `jolt'\ co-vector represents a (translational+rotational)
\textit{collision effect} that can cause hard--tissue spinal injuries.

For example, while driving a car, the SE(3)--jerk of the head--neck system
happens every time the driver brakes abruptly. On the other hand, the
SE(3)--jolt means actual impact to the head. Similarly, the whiplash--jerk,
caused by rear--end car collisions, is like a soft version of the high
pitch--jolt caused by the boxing `upper-cut'. Also, violently shaking the
head left--right in the transverse plane is like a soft version of the high
yaw--jolt caused by the boxing `cross-cut'.

\subsection{Localized spinal dislocations and disclinations caused by the $%
SE(3)-$jolt}

Recall from introduction that for mild (soft--tissue) spinal
injury, the best injury predictor is considered to be the product
of localized spinal strain and strain rate, which is the standard
isotropic viscoelastic continuum concept. To improve this standard
concept, in this subsection, we consider spinal segment (with a
vertebral body, intervertebral disc and other visco-elastic
tissue) as a
3D anisotropic multipolar \emph{Cosserat viscoelastic continuum} \cite%
{Cosserat1,Cosserat2,Eringen02}, exhibiting
coupled--stress--strain elastic properties. This non-standard
continuum model is suitable for analyzing plastic (irreversible)
deformations and fracture mechanics \cite{Bilby} in multi-layered
materials with microstructure (in which slips and bending of
layers introduces additional degrees of freedom, non-existent in
the standard continuum models; see \cite{Mindlin65,Lakes85} for
physical characteristics and \cite{Yang81,Yang82},\\ \cite{Park86}
for biomechanical applications).

The $SE(3)-$jolt $(\mathbf{\dot{F},\dot{T}})$ causes two types of
localized spinal discontinuous deformations:
\begin{enumerate}
\item The Newton jolt $\mathbf{\dot{F}}$ can cause micro-translational \emph{%
dislocations}, or discontinuities in the Cosserat translations;

\item The Euler jolt $\mathbf{\dot{T}}$ can cause micro-rotational \emph{%
disclinations}, or discontinuities in the Cosserat rotations.
\end{enumerate}

For general treatment on dislocations and disclinations related to
asymmetric discontinuous deformations in multipolar materials, see, e.g.,
\cite{Jian95,Yang01}.

To precisely define localized spinal dislocations and disclinations, caused
by the $SE(3)-$jolt $(\mathbf{\dot{F},\dot{T}})$, we first define the
coordinate co-frame, i.e., the set of basis 1--forms $\{dx^{i}\}$, given in
local coordinates $x^{i}=(x^{1},x^{2},x^{3})=(x,y,z)$, attached to spinal
segment's center-of-mass. Then, in the coordinate co-frame $\{dx^{i}\}$ we
introduce the following set of spinal segment's
plastic--deformation--related $SE(3)-$based differential $p-$forms (see \cite%
{14,17}):\newline
$~~~~$the \emph{dislocation current }1--form, $\mathbf{J}=J_{i}\,dx^{i};$%
\newline
$~~~~$the \emph{dislocation density }2--form, $\mathbf{\alpha }=\frac{1}{2}%
\alpha _{ij}\,dx^{i}\wedge dx^{j};$\newline
$~~~~$the \emph{disclination current }2--form, $\mathbf{S}=\frac{1}{2}%
S_{ij}\,dx^{i}\wedge dx^{j};$ ~and\newline
$~~~~$the \emph{disclination density }3--form, $\mathbf{Q}=\frac{1}{3!}%
Q_{ijk}\,dx^{i}\wedge dx^{j}\wedge dx^{k}$,

\noindent where $\wedge $ denotes the exterior wedge--product.
According to Edelen \cite{Edelen,Kadic}, these four $SE(3)-$based
differential forms satisfy the following set of continuity
equations:
\begin{eqnarray}
&&\mathbf{\dot{\alpha}}=\mathbf{-dJ-S,}  \label{dis1} \\
&&\mathbf{\dot{Q}}=\mathbf{-dS,}  \label{dis2} \\
&&\mathbf{d\alpha }=\mathbf{Q,}  \label{dis3} \\
&&\mathbf{dQ}=\mathbf{0,}\qquad  \label{dis4}
\end{eqnarray}
where $\mathbf{d}$ denotes the exterior derivative.

In components, the simplest, fourth equation (\ref{dis4}), representing the
\emph{Bianchi identity}, can be rewritten as
\begin{equation*}
\mathbf{dQ}=\partial _{l}Q_{[ijk]}\,dx^{l}\wedge dx^{i}\wedge dx^{j}\wedge
dx^{k}=0,
\end{equation*}
where $\partial _{i}\equiv\partial /\partial x^{i}$, while $\theta _{\lbrack
ij...]}$ denotes the skew-symmetric part of $\theta _{ij...}$.

Similarly, the third equation (\ref{dis3}) in components reads
\begin{eqnarray*}
\frac{1}{3!}Q_{ijk}\,dx^{i}\wedge dx^{j}\wedge dx^{k} &=&\partial _{k}\alpha
_{\lbrack ij]}\,dx^{k}\wedge dx^{i}\wedge dx^{j},\text{\qquad or} \\
Q_{ijk} &=&-6\partial _{k}\alpha _{\lbrack ij]}.
\end{eqnarray*}

The second equation (\ref{dis2}) in components reads
\begin{eqnarray*}
\frac{1}{3!}\dot{Q}_{ijk}\,dx^{i}\wedge dx^{j}\wedge dx^{k} &=&-\partial
_{k}S_{[ij]}\,dx^{k}\wedge dx^{i}\wedge dx^{j},\text{\qquad or} \\
\dot{Q}_{ijk} &=&6\partial _{k}S_{[ij]}.
\end{eqnarray*}

Finally, the first equation (\ref{dis1}) in components reads
\begin{eqnarray*}
\frac{1}{2}\dot{\alpha}_{ij}\,dx^{i}\wedge dx^{j} &=&(\partial _{j}J_{i}-%
\frac{1}{2}S_{ij})\,dx^{i}\wedge dx^{j},\text{\qquad or} \\
\dot{\alpha}_{ij}\, &=&2\partial _{j}J_{i}-S_{ij}\,.
\end{eqnarray*}

In words, we have:

\begin{itemize}
\item The 2--form equation (\ref{dis1}) defines the time derivative $\mathbf{%
\dot{\alpha}=}\frac{1}{2}\dot{\alpha}_{ij}\,dx^{i}\wedge dx^{j}$ of the
dislocation density $\mathbf{\alpha }$ as the (negative) sum of the
disclination current $\mathbf{S}$ and the curl of the dislocation current $%
\mathbf{J}$.

\item The 3--form equation (\ref{dis2}) states that the time derivative $%
\mathbf{\dot{Q}=}\frac{1}{3!}\dot{Q}_{ijk}\,dx^{i}\wedge dx^{j}\wedge dx^{k}$
of the disclination density $\mathbf{Q}$ is the (negative) divergence of the
disclination current $\mathbf{S}$.

\item The 3--form equation (\ref{dis3}) defines the disclination density $%
\mathbf{Q}$ as the divergence of the dislocation density $\mathbf{\alpha }$,
that is, $\mathbf{Q}$ is the \emph{exact} 3--form.

\item The Bianchi identity (\ref{dis4}) follows from equation
(\ref{dis3}) by \textit{Poincar\'{e} lemma} \cite{14,17} and
states that the disclination density $ \mathbf{Q}$ is conserved
quantity, that is, $\mathbf{Q}$ is the \emph{closed} 3--form.
Also, every 4--form in 3D space is zero.
\end{itemize}

From these equations, we can conclude that localized spinal
dislocations and disclinations are mutually coupled by the
underlaying $SE(3)-$group, which means that we cannot separately
analyze translational and rotational spinal injuries --- a fact
which \emph{is not} supported by the literature.

\section{Conclusion}

Based on the previously developed covariant force law, in this
paper we have formulated a new coupled loading--rate hypothesis,
which states that the main cause of localized spinal injury is an
external $SE(3)-$jolt, an impulsive loading striking the spinal
segment in several degrees-of-freedom, both rotational and
translational, combined. To demonstrate this, we have developed
the vector Newton--Euler mechanics on the Euclidean $SE(3)-$group
of localized spinal micro-motions. In this way, we have precisely
defined the concept of the $SE(3)-$jolt, which is a cause of rapid
localized spinal discontinuous deformations: (i) mild rotational
disclinations and (ii) severe translational dislocations. Based on
the presented model, we argue that we cannot separately analyze
localized spinal rotations from translations, as they are in
reality coupled. To prevent spinal injuries we need to develop the
\textit{internal SE(3)--jolt awareness}. To maintain a healthy
spine, we need to prevent localized SE(3)--jolts from striking any
part of the spine in any human--motion or car--crash conditions.

\section{Appendix: The $SE(3)-$group}

Special Euclidean group $SE(3):=SO(3)\rhd \mathbb{R}^{3}$, (the semidirect
product of the group of rotations with the corresponding group of
translations), is the Lie group consisting of isometries of the Euclidean 3D
space $\mathbb{R}^{3}$.

An element of $SE(3)$ is a pair $(A,a)$ where $A\in SO(3)$ and $a\in \mathbb{%
R}^{3}.$ The action of $SE(3)$ on $\mathbb{R}^{3}$ is the rotation $A$
followed by translation by the vector $a$ and has the expression
\begin{equation*}
(A,a)\cdot x=Ax+a.
\end{equation*}

The Lie algebra of the Euclidean group $SE(3)$ is $\mathfrak{se}(3)=\mathbb{R%
}^{3}\times \mathbb{R}^{3}$ with the Lie bracket
\begin{equation}
\lbrack (\xi ,u),(\eta ,v)]=(\xi \times \eta ,\xi \times v-\eta \times u).
\label{lbse3}
\end{equation}

Using homogeneous coordinates, we can represent $SE(3)$ as follows,
\begin{equation*}
SE(3)=\ \ \left\{ \left(
\begin{array}{cc}
R & p \\
0 & 1%
\end{array}
\right) \in GL(4,\mathbb{R}):R\in SO(3),\,p\in \mathbb{R}^{3}\right\} ,
\end{equation*}
with the action on $\mathbb{R}^{3}$ given by the usual matrix--vector
product when we identify $\mathbb{R}^{3}$ with the section $\mathbb{R}%
^{3}\times \{1\}\subset \mathbb{R}^{4}$. In particular, given
\begin{equation*}
g=\left(
\begin{array}{cc}
R & p \\
0 & 1%
\end{array}
\right) \in SE(3),
\end{equation*}
and $q\in \mathbb{R}^{3}$, we have
\begin{equation*}
g\cdot q=Rq+p,
\end{equation*}
or as a matrix--vector product,
\begin{equation*}
\left(
\begin{array}{cc}
R & p \\
0 & 1%
\end{array}
\right) \left(
\begin{array}{c}
q \\
1%
\end{array}
\right) =\left(
\begin{array}{c}
Rq+p \\
1%
\end{array}
\right) .
\end{equation*}

The Lie algebra of $SE(3)$, denoted $\mathfrak{se}(3)$, is given by \
\begin{equation*}
\mathfrak{se}(3)=\ \ \left\{ \left(
\begin{array}{cc}
\omega & v \\
0 & 0%
\end{array}
\right) \in M_{4}(\mathbb{R}):\omega\in \mathfrak{so}(3),\,v\in \mathbb{R}%
^{3}\right\} ,
\end{equation*}
where the attitude (or, angular velocity) matrix $\omega:\mathbb{R}%
^{3}\rightarrow \mathfrak{so}(3)$ is given by
\begin{equation*}
\omega=\left(
\begin{array}{ccc}
0 & -\omega _{z} & \omega _{y} \\
\omega _{z} & 0 & -\omega _{x} \\
-\omega _{y} & \omega _{x} & 0%
\end{array}
\right) .
\end{equation*}

The \emph{exponential map}, $\exp :\mathfrak{se}(3)\rightarrow SE(3)$, is
given by
\begin{equation*}
\exp \left(
\begin{array}{cc}
\omega & v \\
0 & 0%
\end{array}
\right) =\left(
\begin{array}{cc}
\exp (\omega) & Av \\
0 & 1%
\end{array}
\right) ,
\end{equation*}
where

\begin{equation*}
A=I+\frac{1-\cos \left\Vert \omega \right\Vert }{\left\Vert \omega
\right\Vert ^{2}}\omega+\frac{\left\Vert \omega \right\Vert -\sin \left\Vert
\omega \right\Vert }{\left\Vert \omega \right\Vert ^{3}} \omega^{2},
\end{equation*}
and $\exp (\omega)$ is given by the \emph{Rodriguez' formula},
\begin{equation*}
\exp (\omega)=I+\frac{\sin \left\Vert \omega \right\Vert }{\left\Vert \omega
\right\Vert }\omega+\frac{1-\cos \left\Vert \omega \right\Vert }{\left\Vert
\omega \right\Vert ^{2}}\omega^{2}.
\end{equation*}

\end{document}